\documentclass[12pt]{iopart}
\begin{document}
\jl{1}

\title[Infinite range $t-J$ model]
{Solution of the infinite range $t-J$ model}

\author{B Binz\dag, X Zotos\dag\ddag\ and D Baeriswyl\dag}
\address{\dag\ Institut de Physique th\'eorique, P\'erolles, CH-1700
Fribourg, Switzerland}
\address{\ddag\ Institut Romand de Recherche Num\'erique en Physique des
Mat\'eriaux (IRRMA), EPFL-PPH, CH-1015 Lausanne, Switzerland}

\begin{abstract}
The $t-J$ model with constant $t$ and $J$ between any pair of sites is studied
by exploiting the symmetry of the Hamiltonian with respect to site
permutations. For a given number of electrons and a given total spin the
exchange term simply yields an additive constant. Therefore the real problem is
to diagonalize the "$t$ model", or equivalently the infinite $U$ Hubbard
Hamiltonian. Using extensively the properties of the permutation group, we are
able to find explicitly both the energy eigenvalues and eigenstates, labeled
according to spin quantum numbers and Young diagrams. As a corollary we also
obtain the degenerate ground states of the finite $U$ Hubbard model with
infinite range hopping $-t>0$.
\end{abstract}


\maketitle

\section{Introduction}

It is widely accepted that the $t-J$ model captures the essential physics
of high-temperature superconductors, at least in the normal state
\cite{Dagotto}. The model is defined by the Hamiltonian
\begin{equation}
H = H_t + H_J,
\label{ham}
\end{equation}
where
\begin{equation}
H_t = - \sum_{i,j} t_{ij} P c^{\dagger}_{i \sigma} c_{j \sigma} P
\label{hamt}
\end{equation}
describes the hopping between sites $i$ and $j$, and
\begin{equation}
H_J = - \sum_{i,j} J_{ij} \left(\vec S_i \cdot \vec S_j
- \frac{1}{4} n_i n_j \right)
\label{hamJ}
\end{equation}
is the exchange interaction. The operators $c^{\dagger}_{i \sigma}$
($c_{i \sigma}$) create (destroy) electrons at site $i$ with spin $\sigma$,
$P$ is a projection operator on the subspace with no doubly occupied
sites,
$\vec S_i$ is the spin operator and $n_i$ the particle density restricted
to the values 0 and 1.

Usually both the hopping terms $t_{ij}$ and exchange interactions $J_{ij}$
are chosen to be non-zero if $i$ and $j$ are nearest neighbours and zero
otherwise. Unfortunately, the model is then very hard to solve, and explicit
analytical results have so far only been obtained for a one-dimensional
chain, and even then only for specific values of nearest-neighbour couplings,
namely $J = 2t$ \cite{Schlottmann,Lai} and $J = 0$ \cite{Ogata}.

In this paper we consider the avowedly artificial model with couplings of
unlimited range, i.e. $t_{ij} = t$, $J_{ij} = J$ for all sites i,j. Notice
that the exchange term is then simply given by $-J[S(S+1)-N^2/4]$, where
$N$ is the number of particles. Thus the real problem is to solve the
''$t$ model``, which is equivalent to the infinite $U$ Hubbard model. Models
of this kind have been studied previously \cite{Patterson,vDongen}. A
general solution has been conjectured by Li and Mattis, on the basis of
spectra obtained by exact diagonalization \cite{LiMattis}. Very recently,
Kirson, exploiting the supersymmetry of the model, has calculated
analytically both the energy spectrum and the degeneracies \cite{Kirson}.
One of us (B. B.) has independently solved the model using extensively the
properties of the permutation group \cite{Binz}. This method, described in
detail below, not only offers an alternative way for deriving the energy
eigenvalues but also yields explicitly all the eigenstates.

The paper is organized as follows. Section 2 summarizes the general
properties of the permutation group and its irreducible representations.
 The Young symmetrizers allow to decompose the Hilbert space of
many-electron states into subspaces which transform according to the
irreducible representations of the permutation group. In Section 3 these
symmetrized states are constructed explicitly and characterized by Young
tableaux where the numbers of sites are replaced by symbols indicating the
occupancy of the sites, i.e. $0, \uparrow, \downarrow$. In Section 4 the
Hamiltonian is diagonalized for the subspaces belonging to the different
irreducible representations (or Young diagrams). In Section 5 the technique
is extended to the case of the
Hubbard model (with hopping of unlimited range). This case is in general
more complicated, but for $t<0$ and $U>0$ the exact ground state can be given.
Certain mathematical details are treated in two appendices.

\section{The permutation symmetry and its implications}

The Hilbert space of quantum states $\mathcal{H}$ is generated by the Fock
states
\begin{equation}
|\phi\rangle= c^{\dagger}_{i_1\uparrow}\cdots c^{\dagger}_{i_u\uparrow}
c^{\dagger}_{j_1\downarrow}\cdots c^{\dagger}_{j_d\downarrow}|0\rangle,
\label{conf}
\end{equation}
$u$ and $d$ being the number of $\uparrow$ and $\downarrow$ spins,
respectively. We consider these two quantities as arbitrary but fixed.
We therefore specify the number of electrons $N=u+d$ and the $z-$component
of the total spin $S^z=(u-d)/2$. The number of
empty sites $h$ (called holes) is also conserved with value $h=L-N$,
as doubly occupied states have been excluded.

Since there is a constant hopping amplitude between every pair of sites,
the Hamiltonian is invariant with respect to every
permutation of the lattice sites. The action of such a permutation
$\pi\in S_L$ on a Fock state (\ref{conf}) is given by
the unitary operator $\rho(\pi)$ defined as follows,
\begin{equation}
\rho(\pi) c^{\dagger}_{i_1\sigma_1}\cdots c^{\dagger}_{i_N\sigma_N}|0\rangle:=
c^{\dagger}_{\pi(i_1)\sigma_1}\cdots c^{\dagger}_{\pi(i_N)\sigma_N}|0\rangle.
\label{rep}
\end{equation}
Note that a transposition of two sites occupied by electrons with
the same spin changes the sign of the state whereas the
transposition of empty states leaves it unchanged.

Our approach is based on the commutation relations
\[
[H_t,\rho(\pi)]=0,\ [\vec S,\rho(\pi)]=0 \hbox{  and  }[H_t,\vec S]=0
\hspace{1cm} \forall\pi\in S_L.
\]
They allow us to label each energy level by its total spin quantum
number $S$ and a Young diagram representing the
permutation symmetry. To formulate this more clearly, let us first state
some facts of the representation theory of the
symmetric group developed at different levels in the references
\cite{Messiah,Schensted,JamesKerber}.

There is a one to one correspondence between the irreducible
representations of $S_L$ and the partitions $\alpha$ of $L$
(i.e. the lists
$\alpha=(\alpha_1,\alpha_2,\ldots)$ of integers with the constraints
$\alpha_1+\alpha_2+\ldots=L$ and $\alpha_1\geq\alpha_2\geq\cdots\geq 0$).

The partitions and the corresponding irreducible representations are usually
visualized in terms of a {\it Young diagram} noted $[\alpha]$.
\[
\begin{picture}(150,50)(-10,10)
\put(10,32.5){\normalsize $[\alpha]=$}
\put(40,40){\framebox(10,10){}}
\put(50,40){\framebox(10,10){}}
\put(60,40){\framebox(30,10){\small $\cdots\cdots$}}
\put(90,40){\framebox(10,10){}}
\put(110,42.5){\small $\alpha_1$ boxes}
\put(40,30){\framebox(10,10){}}
\put(50,30){\framebox(10,10){}}
\put(60,30){\framebox(15,10){\small $\cdots$}}
\put(75,30){\framebox(10,10){}}
\put(110,32.5){\small $\alpha_2$ boxes}
\put(60,15){\small $\vdots$}\put(120,15){\small $\vdots$}
\end{picture}
\]
We enumerate the boxes of the diagram column by column from top to bottom
to obtain a {\it Young tableau} $t^\alpha$.
For example, here is the tableau of the diagram $[3,2,1^2]$:
\[
\begin{picture}(80,60)(0,-10)
\put(0,22.5){\normalsize $t^{(3,2,1^2)}=$}
\put(50,30){\framebox(10,10){\small$1$}}
\put(60,30){\framebox(10,10){\small$5$}}
\put(70,30){\framebox(10,10){\small$7$}}
\put(50,20){\framebox(10,10){\small$2$}}
\put(60,20){\framebox(10,10){\small$6$}}
\put(50,10){\framebox(10,10){\small$3$}}
\put(50,0){\framebox(10,10){\small$4$}}
\end{picture}
\]
This tableau defines by its rows the dissection of $\{1,\ldots,7\}$ into
subsets $\{1,5,7\}$, $\{2,6\}$, $\{3\}$ and $\{4\}$,
while it defines by its columns the dissection of $\{1,\ldots,7\}$ into
subsets $\{1,2,3,4\}$, $\{5,6\}$ and $\{7\}$.
Correspondingly we associate with the rows of $t^{(3,2,1^2)}$ the subgroup
$R^{(3,2,1^2)}=S_{\{1,5,7\}}\times S_{\{2,6\}}$
of $S_L$, called  horizontal group or group of the row permutations,
while we obtain from the columns the subgroup
$C^{(3,2,1^2)}=S_{\{1,2,3,4\}}\times S_{\{5,6\}}$ of $S_L$, called vertical
group or group of the  column permutations.
The generalization to arbitrary tableaux is obvious.

We can now define the row symmetrizer
$\mathcal{R}^\alpha:= \sum_{\pi\in R^\alpha}\rho(\pi)$
as well as the  column anti-symmetrizer
$\mathcal{C}^\alpha:=\sum_{\pi\in C^\alpha}{\rm sign}(\pi)\rho(\pi)$.
Finally the Young symmetrizer is given by the product of the two:
$e^\alpha:=\mathcal{C}^\alpha\mathcal{R}^\alpha$.
The Young symmetrizers provide our main working tool for
finding the eigenstates of the model. The following
results of the theory are crucial \cite{Messiah}:

\begin{itemize}
\item {\it Proposition 1:}
If $|\phi\rangle$ is an arbitrary element of $\mathcal{H}$ then
$e^\alpha|\phi\rangle$, if not null, transforms under $S_L$
according to the irreducible representaton $[\alpha]$.

\item {\it Proposition 2:}
For a given Young symmetrizer $e^\alpha$ the set of symmetrized wave functions
$e^\alpha|\phi\rangle$, $|\phi\rangle$  being of the form (\ref{conf}),
spans a subspace $e^\alpha\mathcal{H}$ with a dimension equal to the
number $n_\alpha$ of components $[\alpha]$ contained in $\rho$.\\
Let $|{\Psi_i}\rangle,\ i=1,\ldots,n_\alpha$ be an orthonormal basis of
$e^\alpha\mathcal{H}$, then the space spanned by the vectors
$\rho(\pi)|{\Psi_i}\rangle,\ \pi\in S_L$ ($i$ fixed) is a representation
space for the representation $[\alpha]$ and one obtains $n_\alpha$ mutually
orthogonal representation spaces according to the $n_\alpha$ basis vectors.
\end{itemize}

The problem of diagonalizing $H_t$ and $\vec S^2$ in $\mathcal{H}$ is therefore
completely solved once we have diagonalized it in each of
the subspaces $e^\alpha\mathcal{H}$.
As a final remark let us state that one has also the choice of interchanging
the two factors in the definition of $e^\alpha$
in order to obtain $\tilde e^\alpha=\mathcal{R}^\alpha\mathcal{C}^\alpha$.
Propositions 1 and 2 are true for $\tilde e^\alpha$ as well as for
$e^\alpha$ and we are free to work with either of them.

\section{Construction of symmetrized states}\label{method}

We will now apply the Young symmetrizer of a given tableau to the different
Fock states in order to obtain symmetrized
wavefunctions $e^\alpha|\phi\rangle$. Such a state is best represented
graphically in terms of the corresponding tableau, where we
replace the number $i\in\{1,\ldots,L\}$ of each box by the occupancy
$\uparrow$, $\downarrow$ or $0$ (empty) of the lattice site $i$ in
$|\phi\rangle$.

For example in a system of $L=7$ sites with two up and two down
spins the tableau $t^{(3,2,1^2)}$
yields the following symmetrized wave functions:
\begin{equation}
\begin{picture}(130,110)(0,-20)
\put(0,80){\framebox(10,10){\small$0$}}
\put(10,80){\framebox(10,10){\small$\uparrow$}}
\put(20,80){\framebox(10,10){\small$\downarrow$}}
\put(0,70){\framebox(10,10){\small$0$}}
\put(10,70){\framebox(10,10){\small$\downarrow$}}
\put(0,60){\framebox(10,10){\small$0$}}
\put(0,50){\framebox(10,10){\small$\uparrow$}}
\put(40,72.5){\normalsize$=e^{(3,2,1^2)}c^{\dagger}_{4\uparrow}
c^{\dagger}_{5\uparrow}c^{\dagger}_{6\downarrow}
c^{\dagger}_{7\downarrow}|0\rangle$}
\put(0,30){\framebox(10,10){\small$0$}}
\put(10,30){\framebox(10,10){\small$0$}}
\put(20,30){\framebox(10,10){\small$0$}}
\put(0,20){\framebox(10,10){\small$\uparrow$}}
\put(10,20){\framebox(10,10){\small$\uparrow$}}
\put(0,10){\framebox(10,10){\small$\downarrow$}}
\put(0,0){\framebox(10,10){\small$\downarrow$}}
\put(40,22.5){\normalsize
$=e^{(3,2,1^2)}c^{\dagger}_{2\uparrow}c^{\dagger}_{6\uparrow}
c^{\dagger}_{3\downarrow}c^{\dagger}_{4\downarrow}|0\rangle$}
\put(40,-10){\normalsize\vdots}
\end{picture}
\label{expl2}
\end{equation}
The question is now: how many (and which) of the
${7!\over2!\cdot2!\cdot3!}=210$ states given above are linearly independent?

Indeed there is a way to answer this question without doing explicit
calculations. First, according to the definition of $e^\alpha$, for two
configurations wich differ only by a row permutation the results of the
symmetrization are identical ($|\phi\rangle =\rho(\pi)|{\phi'}\rangle,
\ \pi\in R^\alpha \Longrightarrow
e^\alpha|\phi\rangle=e^\alpha|{\phi'}\rangle$).
It implies that $e^\alpha |\phi\rangle$ is zero whenever two equally
oriented spins are in the same row of the corresponding tableau,
as in the second row of (\ref{expl2}). This observation can be
converted into a graphical rule that eliminates vanishing or linearly
dependent states: Choose an order in the three symbols
$\uparrow,\downarrow,0$ e.g. $0<\uparrow<\downarrow$ and take only
into account the graphs, whose rows are filled in non-decreasing order;
in addition make sure that there be no repeated $\uparrow$ or $\downarrow$
symbols in the rows.
If we had worked with $\tilde e^\alpha$ instead of $e^\alpha$, we would
find another rule, which this time involves the columns of a tableau
instead of the rows and the holes instead of the spins: The rule states
that there cannot be two holes in the same column.

It seems then natural (although not immediately obvious) to merge these
two rules into a single statement:

\begin{itemize}
\item {\it Proposition 3:}
A basis of the subspace $e^\alpha\mathcal{H}$ is given by the symmetrized
wavefunctions $e^\alpha|\phi\rangle$ whose graphical representations
obey the following conditions:
\begin{enumerate}
\item The rows from left to right and the columns from top to bottom are
filled in a non-decreasing order with the symbols $0<\uparrow<\downarrow$.
\item No two equally oriented spins are in the same row.
\item No two holes are in the same column.
\end{enumerate}
\end{itemize}

As a corrollary, the multiplicity $n_\alpha$ of the irreducible
representation $[\alpha]$ in $\rho$ equals the number of admissible ways
 of filling the diagram $[\alpha]$ with the symbols
$\underbrace{0,\ldots,0}_{h\times},
\underbrace{\uparrow,\ldots,\uparrow}_{u\times},
\underbrace{\downarrow,\ldots,\downarrow}_{d\times}$
according to these conditions.

Although proposition 3 is simple, reflecting in a natural way the fermionic
nature of electrons and the bosonic nature of
holes, it is not easy to prove it directly. It can nevertheless be seen
to be a special case of the Littlewood-Richardson rule,
as explained in appendix A.

\section{Spectrum and eigenstates of the model}

A simple example of diagonalization using the Young symmetrizers is
the single-particle problem. For $N=1$ and $L-1$ empty
sites one can build two distinct tableaux:
\[
\begin{picture}(105,20)
\put(0,10){\framebox(10,10){\small $0$}}
\put(10,10){\framebox(15,10){\small $\cdots$}}
\put(25,10){\framebox(10,10){\small $0$}}
\put(35,10){\framebox(10,10){\small $\uparrow$}}
\put(50,10){,}\put(65,10){\framebox(10,10){\small $0$}}
\put(75,10){\framebox(15,10){\small $\cdots$}}
\put(90,10){\framebox(10,10){\small $0$}}
\put(65,0){\framebox(10,10){\small $\uparrow$}}\put(105,10){.}
\end{picture}
\]
The former corresponds to the nondegenerate eigenstate
$c^+_{0\uparrow}|0\rangle:=\sum_i c^+_{i\uparrow}|0\rangle$ with energy
$E=-Lt$ and the latter to the eigenstate
$(c^+_{2\uparrow}-c^+_{1\uparrow})|0\rangle$
which is $L-1$ fold degenerate with energy $E=0$.

A closer view on proposition 3 shows that the most general allowed
diagram in the many-particle problem is of the form
$[\alpha]=[l,2^{m-1},1^{k-m}]$ with the associated tableau:

\begin{equation}
\begin{picture}(200,170)(-50,-125)
\put(-50,-35){\normalsize $t^\alpha=$}
\put(55,35){\normalsize$l$}
\put(1,30){\line(1,0){108}}
\put(0,27.5){\small$\leftarrow$}
\put(100,27.5){\small$\rightarrow$}
\put(0,0){\framebox(20,20){\normalsize$1$}}
\put(20,0){\framebox(20,20){{\tiny $k+1$}}}
\put(40,0){\framebox(20,20){
{\tiny $\hspace{-2pt}\begin{array}{c}k+m\\ +1\end{array}$}}}
\put(60,0){\framebox(30,20){\normalsize$\cdots$}}
\put(90,0){\framebox(20,20){\normalsize$L$}}
\put(0,-30){\framebox(20,30){\normalsize$\vspace{5pt}\vdots$}}
\put(20,-30){\framebox(20,30){\normalsize$\vspace{5pt}\vdots$}}
\put(0,-50){\framebox(20,20){\normalsize$m$}}
\put(20,-50){\framebox(20,20){{\tiny $k+m$}}}
\put(0,-70){\framebox(20,20){{\tiny $m+1$}}}
\put(0,-100){\framebox(20,30){\normalsize$\vspace{5pt}\vdots$}}
\put(0,-120){\framebox(20,20){\normalsize$k$}}
\put(-10,19){\line(0,-1){138}}
\put(-12.5,-117.5){\small$\downarrow$}
\put(-12.5,12.5){\small$\uparrow$}
\put(-20,-50){\normalsize$k$}
\put(120,19){\line(0,-1){68}}
\put(117.5,-47.5){\small$\downarrow$}
\put(117.5,12.5){\small$\uparrow$}
\put(130,-15){\normalsize$m$}
\end{picture}
\label{klm}
\end{equation}

The width $l$ of the first row is restricted to the values $L-N, L-N+1$
and $L-N+2$, whereas the allowed $k$ and $m$
values depend on $S^z$ (the numbers of $\uparrow$ and $\downarrow$ electrons).

Once the relevant irreducible representations are specified, one can
diagonalize $H_t$ within the subspaces
$e^\alpha\mathcal{H}$. For this purpose it is extremely convenient that $H_t$
can be expressed in terms of permutation operators. One finds
\[
H_t=-t\left[ \sum_{i<j}\rho((ij))+\vec S^2+f(L,N)\right],
\]
with $f(L,N)=N^2/4-(L-N)(L-N-1)/2$ and $(ij)$ the transposition of
sites $i$ and $j$.

The energy of an eigenstate is thus completely determined by its
symmetry $[\alpha]$ and its total spin $S$.
With the aid of the algebraic lemma
\begin{equation}
\fl\sum_{i<j}\rho((ij))e^\alpha=( {\hbox{ \# transpositions in }
R^\alpha-\hbox{ \# transpositions in }C^\alpha})\ e^\alpha
\label{lemma}
\end{equation}
(proved in appendix B), we compute the energy as a function of $S$
and $[\alpha]=[l,2^{m-1},1^{k-m}]$. In this way we obtain the complete spectrum
 of the Hamiltonian $H_t$,
\begin{eqnarray}
E(S,\alpha)&=&-t\left[\left(\begin{array}{c}l\\ 2\end{array}\right)-
\left(\begin{array}{c}k\\ 2\end{array}\right)-
\left(\begin{array}{c}m-1\\ 2\end{array}\right)\right.\nonumber\\
& &+\left.\left(\begin{array}{c}\frac N2+S+1\\ 2\end{array}\right)+
\left(\begin{array}{c}\frac N2-S\\ 2\end{array}\right)-
\left(\begin{array}{c}L-N\\ 2\end{array}\right)\right].\label{energy}
\end{eqnarray}

The spectrum is shown in figure \ref{sp1} (\ref{sp2}) for an even (odd)
number of electrons $1<N<L-1$
(the case $N=L-1$ with only one hole is treated separately).
We have assumed that $t$ is positive. In the opposite case the spectrum
is simply inverted.
Apart from the energy values we indicate also the total spin $S$.
The column on the right-hand side of the figures refers to
the permutation symmetry.

Due to the large symmetry group every single energy level of this system
will in general be highly degenerate.
The degeneracy of a level corresponding to $[\alpha]$ and $S$ is $(2S+1)$
times the degree $f^\alpha$ of the irreducible
representation $[\alpha]$. The latter can be calculated following the
references \cite{Schensted} or \cite{JamesKerber}
and in our case amounts to
\begin{equation}
f^\alpha={L!\over k!\,(m-1)!\,(l-2)!}\cdot{k-m+1\over(k+l-1)(m+l-2)}.
\label{edeg}
\end{equation}

We distinguish four parts labeled by capital letters, that we will now
discuss seperately:

\subsection{A: $l=L-N+2$ and B: $l=L-N$}

Consider all the Young diagrams with $l=L-N+2$ (case A) or $l=L-N$ (case B).
In both cases the multiplicity
is $n_\alpha=1$ if $|S^z|\leq(k-m)/2$ and $0$ elsewhere. To see this, we
look at the tableaux in the equations (\ref{caseA}) and (\ref{caseB}) which
represent the only allowed filling according to proposition 3. One has the
liberty to invert some of the $\uparrow$ spins in the $k-m$ last boxes of the
first column, but not more than these. Hence the total spin is
$S=(k-m)/2$. By varying $k$ and $m$ with $l$ and $L$ fixed one obtains
every possible value for $S$ in the case B and every value exept the
completely magnetized $S=N/2$ in the case A.

The energies given by (\ref{energy}) turn out to be $E_A=-t(2L-N)$
(resp. $E_B=0$) independently of the different values of
$S$, which leads to an accidental degeneracy. This means that there
are states of different symmetries and spin values
with the same energy. This degeneracy is lifted by the term $H_J$
in the Hamiltonian (\ref{ham}).

Because the dimension of $e^\alpha\mathcal{H}$ is one,
the state $e^\alpha|\phi\rangle$
is an eigenstate of $H_t$.
In the case A ($l=L-n+2$), it is convenient to use $\tilde e^\alpha$
instead of $e^\alpha$ and to change
the order convention of proposition 3 into $\uparrow<\downarrow<0$.
We then obtain the eigenstates A:
\begin{equation}
\fl\begin{picture}(250,100)(0,-92.5)
\put(0,-2.5){\framebox(10,10){\small$\uparrow$}}
\put(10,-2.5){\framebox(10,10){\small$\downarrow$}}
\put(20,-2.5){\framebox(10,10){\small$0$}}
\put(30,-2.5){\framebox(15,10){\small$\cdots$}}
\put(45,-2.5){\framebox(10,10){\small$0$}}
\put(0,-12.5){\framebox(10,10){\small$\uparrow$}}
\put(10,-12.5){\framebox(10,10){\small$\downarrow$}}
\put(0,-27.5){\framebox(10,15){\small$\vspace{5pt}\vdots$}}
\put(10,-27.5){\framebox(10,15){\small$\vspace{5pt}\vdots$}}
\put(0,-37.5){\framebox(10,10){\small$\uparrow$}}
\put(10,-37.5){\framebox(10,10){\small$\downarrow$}}
\put(0,-47.5){\framebox(10,10){\small$\uparrow$}}
\put(0,-62.5){\framebox(10,15){\small$\vspace{5pt}\vdots$}}
\put(0,-72.5){\framebox(10,10){\small$\uparrow$}}
\put(65,-20){ \normalsize $
=b^{\dagger}_{2\,k+2}\cdots b^{\dagger}_{m\,k+m}c^{\dagger}_{m+1\,\uparrow}
\cdots c^{\dagger}_{k\,\uparrow}
\sum_{i<j\in\Lambda}b^{\dagger}_{ij}|0\rangle, $}
\end{picture}\label{caseA}
\end{equation}
where $b^{\dagger}_{ij}:=c^{\dagger}_{i\uparrow}c^{\dagger}_{j\downarrow}
+c^{\dagger}_{j\uparrow}c^{\dagger}_{i\downarrow}$ creates a singlet
pair on the sites $i$ and $j$ and $\Lambda$ is the sublattice
formed by the $l$ sites in the first row of $t^\alpha$. This is only one
particular eigenstate of this level. In fact, as already stated in
reference \cite{vDongen}, a general eigenstate of level A
with total spin $S$ is of the form:
\[
P\psi^{\dagger}_{(N-2,S)}c^+_{0\uparrow}c^+_{0\downarrow}|0\rangle,
\]
where $\psi^{\dagger}_{(N-2,S)}$ is an arbitrary wavefunction of $N-2$
electrons with spin $S$, $c^+_{0\sigma}:=\sum_i c^+_{i\sigma}$ creates an
electron in the single-particle groundstate and $P$ projects out states with
doubly occupied sites.

An eigenstate of level B is given by
\begin{equation}
\fl\begin{picture}(250,100)(0,-92.5)
\put(0,-2.5){\framebox(10,10){$0$}}
\put(10,-2.5){\framebox(10,10){$0$}}
\put(20,-2.5){\framebox(10,10){$0$}}
\put(30,-2.5){\framebox(15,10){$\cdots$}}
\put(45,-2.5){\framebox(10,10){$0$}}
\put(0,-12.5){\framebox(10,10){$\uparrow$}}
\put(10,-12.5){\framebox(10,10){$\downarrow$}}
\put(0,-27.5){\framebox(10,15){$\vspace{5pt}\vdots$}}
\put(10,-27.5){\framebox(10,15){$\vspace{5pt}\vdots$}}
\put(0,-37.5){\framebox(10,10){$\uparrow$}}
\put(10,-37.5){\framebox(10,10){$\downarrow$}}
\put(0,-47.5){\framebox(10,10){$\uparrow$}}
\put(0,-62.5){\framebox(10,15){$\vspace{5pt}\vdots$}}
\put(0,-72.5){\framebox(10,10){$\uparrow$}}
\put(65,-20){ \normalsize $
=\sum_{\pi\in S_k}\sum_{\tau\in S_m}{\rm sign}(\pi\tau)
b^{\dagger}_{\pi(2)\tau(k+2)}\cdots b^{\dagger}_{\pi(m)\tau(k+m)}
c^{\dagger}_{\pi(m+1)\uparrow}
\cdots c^{\dagger}_{\pi(k)\uparrow} |0\rangle $ ,}
\end{picture}
\label{caseB}
\end{equation}
$S_k$ being the set of permutations of $\{1,\ldots,k\}$ (the first column)
and $S_m$ the permutations of
$\{k+1,\ldots,k+m\}$ (second column). Again this is only one representative
member of a large subspace of degenerate eigenstates.
The others can in principle be calculated through the repeated application
of permutation operators, the spin-lowering
operator $S^-:=S^x-iS^y$ and linear combinations of them. Whether there
is a more compact characterization of these subspaces
like in the case A is an open question.

\subsection{C and D: $l=L-N+1$}

The diagrams with $l=L-n+1$ appear with multiplicity $2$ if
$|S^z|\leq(k-m-1)/2$ and with multiplicity
$1$ if $|S^z|=(k-m+1)/2$. The diagonalization of $H_t$ in $e^\alpha\mathcal{H}$
leads therefore to two levels with total spin $S=(k-m\pm1)/2$.
$e^\alpha\mathcal{H}$ is spanned by the two symmetrized wavefunctions:
\[
\begin{picture}(230,110)(0,-92.5)
\put(0,-2.5){\framebox(10,10){\small$0$}}
\put(10,-2.5){\framebox(10,10){\small$0$}}
\put(20,-2.5){\framebox(15,10){\small$\cdots$}}
\put(35,-2.5){\framebox(10,10){\small$0$}}
\put(45,-2.5){\framebox(10,10){\small$\uparrow$}}
\put(0,-12.5){\framebox(10,10){\small$\uparrow$}}
\put(10,-12.5){\framebox(10,10){\small$\downarrow$}}
\put(0,-27.5){\framebox(10,15){\small$\vspace{5pt}\vdots$}}
\put(10,-27.5){\framebox(10,15){\small$\vspace{5pt}\vdots$}}
\put(0,-37.5){\framebox(10,10){\small$\uparrow$}}
\put(10,-37.5){\framebox(10,10){\small$\downarrow$}}
\put(0,-47.5){\framebox(10,10){\small$\uparrow$}}
\put(0,-62.5){\framebox(10,15){\small$\vspace{5pt}\vdots$}}
\put(0,-72.5){\framebox(10,10){\small$\uparrow$}}
\put(0,-82.5){\framebox(10,10){\small$\downarrow$}}
\put(65,-20){\normalsize$=:|{\Psi_1}\rangle$,}

\put(125,-2.5){\framebox(10,10){\small$0$}}
\put(135,-2.5){\framebox(10,10){\small$0$}}
\put(145,-2.5){\framebox(15,10){\small$\cdots$}}
\put(160,-2.5){\framebox(10,10){\small$0$}}
\put(170,-2.5){\framebox(10,10){\small$\downarrow$}}
\put(125,-12.5){\framebox(10,10){\small$\uparrow$}}
\put(135,-12.5){\framebox(10,10){\small$\downarrow$}}
\put(125,-27.5){\framebox(10,15){\small$\vspace{5pt}\vdots$}}
\put(135,-27.5){\framebox(10,15){\small$\vspace{5pt}\vdots$}}
\put(125,-37.5){\framebox(10,10){\small$\uparrow$}}
\put(135,-37.5){\framebox(10,10){\small$\downarrow$}}
\put(125,-47.5){\framebox(10,10){\small$\uparrow$}}
\put(125,-62.5){\framebox(10,15){\small$\vspace{5pt}\vdots$}}
\put(125,-72.5){\framebox(10,10){\small$\uparrow$}}
\put(125,-82.5){\framebox(10,10){\small$\uparrow$}}
\put(190,-20){\normalsize$=:|{\Psi_2}\rangle$.}
\end{picture}
\]

The odd combination of them
\begin{eqnarray*}
\fl|{\Psi_2}\rangle-|{\Psi_1}\rangle
&=&\sum_{\pi\in S_k}\sum_{\tau\in S_m}{\rm sign}(\pi\tau)
b^{\dagger}_{\pi(2)\tau(k+2)}\cdots b^{\dagger}_{\pi(m)\tau(k+m)}\\
 & &c^{\dagger}_{\pi(m+1)\uparrow}\cdots c^{\dagger}_{\pi(k-1)\uparrow}
(b^{\dagger}_{\pi(k)\tau(k+1)}+\sum_{\nu =k+m+1}^L
b^{\dagger}_{\pi(k)\,\nu})|0\rangle
\end{eqnarray*}
with $S^z=(k-m-1)/2$
is easily seen to be an eigenstate of $\vec S^2$ with $S=(k-m-1)/2$
because it is annulled by the raising operator
$S^{+}=S^x+iS^y$. Hence it has to be an eigenstate of $H_t$ as well.
The states of this type give rise to the part C
of the spectrum with energies $E_C=-t(L-N/2-1-S)$.

The second eigenvector with $S=(k-m+1)/2$ must be orthogonal to
$|\Psi_2\rangle-|\psi_1\rangle$ and therefore is given by the sum
\begin{eqnarray*}
\fl|{\Psi_2}\rangle+|{\Psi_1}\rangle
&=&\sum_{\pi\in S_k}\sum_{\tau\in S_m}{\rm sign}(\pi\tau)
b^{\dagger}_{\pi(2)\tau(k+2)}\cdots b^{\dagger}_{\pi(m)\tau(k+m)}\\
 & &c^{\dagger}_{\pi(m+1)\uparrow}\cdots c^{\dagger}_{\pi(k-1)\uparrow}
(d^{\dagger}_{\pi(k)\pi(1)}+
d^{\dagger}_{\pi(k)\tau(k+1)}+\sum_{\nu =k+m+1}^L
d^{\dagger}_{\pi(k)\,\nu})|0\rangle,
\end{eqnarray*}
with $d^{\dagger}_{ij}:=c^{\dagger}_{i\uparrow}c^{\dagger}_{j\downarrow}-
c^{\dagger}_{j\uparrow}c^{\dagger}_{i\downarrow}$.
These states correspond to the part D of the spectrum
and the energies are $E_D=-t(L-N/2+S)$.

\subsection{The special case of one single hole ($N=L-1$)}

For $N=L-1$ figures 1 and 2 remain valid exept that level B contains
now only the ferromagnetic states with $S=N/2$. To see this, we notice
that the symmetrized states of level B correspond to a Young diagram with
$l=L-N$. If only one hole is present, this means
that $l=1$ and thus the only remaining diagram is $[1^L]$.
This diagram corresponds to the ferromagnetic state
\[
\sum_{i=1}^L(-1)^ic_{1\uparrow}^{\dagger}\cdots
\widehat{c_{i\uparrow}^{\dagger}}\cdots
c_{L\uparrow}^{\dagger}|0\rangle.
\]
The only degeneracy is in this case the trivial spin degeneracy $2S+1$.

It follows that for $N=L-1$ and a positive hopping parameter $(-t>0)$
the ground state is ferromagnetic.
This result is not surprising since it is a consequence of two  well
known theorems, both confirming
a unique ferromagnetic ground state for this particular case.
The first is Tasaki's extension of Nagaoka's theorem
\cite{Tasaki} and the second is a theorem proven by Mielke on flat band
ferromagnetism \cite{Mielke,Pieri}.

In this model, we find an example of Nagaoka ferromagnetism where the
one-hole-condition is absolutely  necessary, for
we find always a complete spin degeneracy for $N<L-1$.

\subsection{Permutation symmetry and supersymmetry}

At this point it is worthwile to connect the present approach with that of
Kirson \cite{Kirson}, who exploited the dynamical supersymmetry of the model.
His classification of many-electron states in Fock space $\mathcal F$ is based
on the irreducible representations $[Y,S]$ of a certain superalgebra. The
representation space of $[Y,S]$ contains four possible pairs of quantum
numbers, namely $(Y,S)$, $(Y+\case12,S-\case12)$, $(Y-\case12,S-\case12)$ and
$(Y,S-1)$, where $Y=L-\case12 N$ and $S$ is the total spin. We can identify the
representation space of $[Y,S]$ with $e^\alpha{\mathcal F}$, where
$[\alpha]=[l,2^{m-1},1^{k-m}]$ is related to $Y$ and $S$ by $k=L-Y+S$,
$m=L-Y-S+1$ and $l=-L+2Y+1$.

For a given $\alpha$ there are (in general) four classes of symmetrized states
in Fock space. These correspond to parts A to D of the spectrum with differing
numbers of particles ($N$ and $N\pm1$), and can be identified with the four
pairs
of quantum numbers in $[Y,S]$:
\[
{\rm A}\Leftrightarrow (Y-\case12,S-\case12),\ {\rm B}\Leftrightarrow
(Y+\case12,S-\case12),\ {\rm C}\Leftrightarrow (Y,S-1),\ {\rm
D}\Leftrightarrow (Y,S).
\]

\section{Ground states of the finite $U$ Hubbard model with infinite
range hopping in the case $-t>0$}\label{Hubbard}

The method developed in the previous sections can be
generalized for other models which
are invariant under permutations of the lattice sites. For instance the
Hubbard model with infinite range hopping,
\[ H_{\rm Hubb}=H_0+H_U \]
\[ H_0=-t\sum_{i,j,\sigma}c_{i\sigma}^{\dagger}c_{j\sigma}^{} \]
\[ H_U=U\sum_i n_{i\uparrow}n_{i\downarrow} \]
can in principle be treated in the same way. The most important new feature
is the appearance of doubly occupied sites.
Proposition 3 has to be modified in a way as to treat these sites
as well. The procedure is the following:
\begin{enumerate}
\item Compute the admissible tableaux (the basis states of
$e^\alpha\mathcal{H}$) without double occupation as explained in section
\ref{method} or in appendix A.
\item Replace a pair $\uparrow,\downarrow$ of symbols by $\uparrow\downarrow,0$
and compute the symmetrized
states with exactly one double occupation. The new symbol $\uparrow\downarrow$
 has to be included in the ordering convention, e.g.
$0<\uparrow<\downarrow<\uparrow\downarrow$.
\item  Replace another two symbols $\uparrow,\downarrow$ by
$\uparrow\downarrow,0$ and continue, until there is no pair
$\uparrow,\downarrow$ left. In proposition 3, the symbols $\uparrow\downarrow$
are treated like the holes, i.e. they must not be repeated within the same
column.
\end{enumerate}

A model which includes doubly occupied sites is much more difficult to
solve than the model considered in this paper. Nevertheless there is
a particular class of diagrams where these complications do not matter.

Consider a diagram of the form (\ref{klm}) where the number $l$ of boxes
in the first row equals $L-N$.  In the tableaux of this kind, there is no way
to produce a doubly occupied state without violating the rules, because
there is no room for an additional hole.
The only symmetrized states according to such a diagram are therefore
the states B, eigenstates of $H_t$, which contain
no double occupation. We conclude that every eigenstate of $H_t$
belonging to case B is at the same time an eigenstate of the Hubbard
Hamiltonian  with the unchanged energy $E_B=0$.

Since in the case $-t>0$ and $U>0$ we find $\langle\psi|
H_0|\psi\rangle\geq 0$
and $\langle\psi| H_U|\psi\rangle\geq 0$ for every state
$|\psi\rangle$, the states B are even the (only) ground states of
$H_{\rm Hubb}$.
It is remarkable that the term $H_J$ splits the accidental degeneracy of
level B, while this degeneracy remains exact in
the Hubbard model for every positive value of $U$. This shows that the
$t-J$ model does not capture correctly the behaviour of the Hubbard model,
even not in the asymptotic region $U>>|t|$. (In fact, a systematic large $U$
expansion of the Hubbard model yields , in addition to the exchange term,
another contribution, the so-called pair-hopping term. See e.g.
\cite{Auerbach}.)

\section{Conclusion}

We have shown that the permutation symmetry of the $t-J$ model with infinite
range hopping allows to derive explicitly the energy spectrum, the
eigenfunctions and their quantum numbers. The model is admittedly rather
unphysical due to the complete lattice connectivity which leads to unusually
high level degeneracies. Nevertheless the many-body spectrum has a very rich
structure, and therefore the model deserves to be added to the few nontrivial
cases of exactly solvable strongly correlated fermion systems. Our results for
the spectrum and the degeneracies agree with those derived on the basis of a
dynamical supersymmetry \cite{Kirson}, but in addition we have also been able
to obtain all the eigenstates. Furthermore, we have found an exact
correspondence between the two approaches.

\ack
We are indebted to Duncan Haldane for initiating this study several years
ago. We are grateful to Pierbiagio Pieri for pointing out to us reference
\cite{Kirson}
and for many valuable discussions which finally led to section \ref{Hubbard}.
We also thank Claude Auderset for indicating to us reference
\cite{JamesKerber}. This work was supported by the Swiss National Foundation
through grant No. 20-46918.96.

\appendix
\section{The Littlewood-Richardson rule and proof of proposition 3}

Sometimes a representation of a group $G$ is completely determined by a
representation of a soubgroup of $G$. To formulate this properly, we refer
to the concept  of induced representations.

\begin{itemize}
\item {Proposition A.1:}
Given a soubgroup $H$ of a finite group $G$ and a representation $\sigma$
of $H$, there exists
always a representation $\rho$ of $G$ into a vector space $V$ and
 a subrepresentation $\tilde\sigma$ of $\rho|_H$
into a subspace $W$ of $V$ such that $\tilde\sigma$ is equivalent to $\sigma$
and
\[ V=\bigoplus_{\gamma\in G/H}\ W_\gamma, \]
where $G/H$ is the set of left cosets of $H$ in $G$ and $W_\gamma=\rho(s)W$
for an arbitrary $s\in \gamma$.
\end{itemize}

In this situation, $\rho$ is (up to an equivalency) uniquely determined
by $\sigma$ and is called the induction of $\sigma$ into $G$.

Let $[\alpha]$ be an irreducible representation of $S_n$ and $[\beta]$ an
irreducible representation of $S_m$, then the
tensor product $[\alpha]\otimes[\beta]$ yields an irreducible representation
of $S_n\times S_m$. $S_n\times S_m$ can
be identified in a natural way with a subgroup of $S_{n+m}$, if $S_n$ acts
on the elements $\{1,2,\ldots,n\}$ and $S_m$
acts on $\{n+1,n+2,\ldots,n+m\}$. The outer product $[\alpha][\beta]$
is defined as the induction of
$[\alpha]\otimes[\beta]$ into $S_{n+m}$ and is in general a reducible
representation of $S_{n+m}$. This multiplication
is associative, commutative and obeys a distributive law together with
the direct sum $\oplus$.

The representation $\rho$ defined in equation (\ref{rep}) is an outer product:
\[ \rho=[h][1^u][1^d]. \]
To see this, consider one particular Fock state $|\phi\rangle$ of the
form (\ref{conf}). The subgroup of $S_L$
that leaves $|\phi\rangle$ invariant (up to a sign) is isomorphic to
$S_h\times S_u\times S_d$.
The one-dimensional subspace $W$ of $\mathcal{H}$ generated by $|\phi\rangle$
carries therefore the representation
$\sigma=[h]\otimes [1^u]\otimes [1^d]$ of $S_h\times S_u\times S_d<S_L$.
All we have to verify is that the Hilbert space of the system
(with $N$ and $S^z$ fixed) is the direct sum
\[ {\mathcal{H}}=\bigoplus_{\gamma\in S_L/{S_h\times S_u\times S_d}}
\rho(\pi_\gamma)\,W, \]
where $\pi_\gamma\in\gamma$ is a representative member of the left coset
$\gamma$.

The Littlewood-Richardson rule describes a graphical way to generate the
irreducible constituents of an arbitrary product $[\alpha][\beta]$,
but for our purpose it is sufficient to consider a product of the form
$[\alpha][1^n]$:

\begin{itemize}
\item {Proposition A.2:}
The diagrams $[\gamma]$ of the irreducible constituents of $[\alpha][1^n]$
may be calculated by adding $n$ boxes to the
diagram $[\alpha]$ in all possible ways such that no two added boxes
appear in the same row.
\end{itemize}

Example: if $[\alpha]=[3,1^2]$ and $n=2$ we obtain:
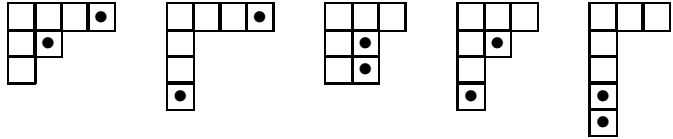
\[
\begin{picture}(250,70)(0,-52.5)
\put(0,-2.5){\framebox(10,10){}}
\put(10,-2.5){\framebox(10,10){}}
\put(20,-2.5){\framebox(10,10){}}
\put(30,-2.5){\framebox(10,10){\small$\bullet$}}
\put(0,-12.5){\framebox(10,10){}}
\put(10,-12.5){\framebox(10,10){\small$\bullet$}}
\put(0,-22.5){\framebox(10,10){}}

\put(60,-2.5){\framebox(10,10){}}
\put(70,-2.5){\framebox(10,10){}}
\put(80,-2.5){\framebox(10,10){}}
\put(90,-2.5){\framebox(10,10){\small$\bullet$}}
\put(60,-12.5){\framebox(10,10){}}
\put(60,-22.5){\framebox(10,10){}}
\put(60,-32.5){\framebox(10,10){\small$\bullet$}}

\put(120,-2.5){\framebox(10,10){}}
\put(130,-2.5){\framebox(10,10){}}
\put(140,-2.5){\framebox(10,10){}}
\put(120,-12.5){\framebox(10,10){}}
\put(130,-12.5){\framebox(10,10){\small$\bullet$}}
\put(120,-22.5){\framebox(10,10){}}
\put(130,-22.5){\framebox(10,10){\small$\bullet$}}

\put(170,-2.5){\framebox(10,10){}}
\put(180,-2.5){\framebox(10,10){}}
\put(190,-2.5){\framebox(10,10){}}
\put(170,-12.5){\framebox(10,10){}}
\put(180,-12.5){\framebox(10,10){\small$\bullet$}}
\put(170,-22.5){\framebox(10,10){}}
\put(170,-32.5){\framebox(10,10){\small$\bullet$}}

\put(220,-2.5){\framebox(10,10){}}
\put(230,-2.5){\framebox(10,10){}}
\put(240,-2.5){\framebox(10,10){}}
\put(220,-12.5){\framebox(10,10){}}
\put(220,-22.5){\framebox(10,10){}}
\put(220,-32.5){\framebox(10,10){\small$\bullet$}}
\put(220,-42.5){\framebox(10,10){\small$\bullet$}}
\end{picture}
\]
Thus
\[ [3,1^2][1^2]=[4,2,1]\oplus [4,1^3] \oplus [3,2^2]
\oplus [3,2,1^2] \oplus [3,1^4]. \]

We can apply this process repeatedly in order to get the
constituents of $\rho=[h][1^u][1^d]$.

For example, if we want to calculate $[3][1^2][1^2]$,
we first evaluate $[3][1^2]$:

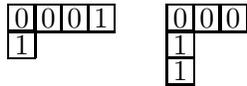
\[
\begin{picture}(90,50)(0,-32.5)
\put(0,-2.5){\framebox(10,10){\small$0$}}
\put(10,-2.5){\framebox(10,10){\small$0$}}
\put(20,-2.5){\framebox(10,10){\small$0$}}
\put(30,-2.5){\framebox(10,10){\small$1$}}
\put(0,-12.5){\framebox(10,10){\small$1$}}

\put(60,-2.5){\framebox(10,10){\small$0$}}
\put(70,-2.5){\framebox(10,10){\small$0$}}
\put(80,-2.5){\framebox(10,10){\small$0$}}
\put(60,-12.5){\framebox(10,10){\small$1$}}
\put(60,-22.5){\framebox(10,10){\small$1$}}
\end{picture}
\]
This yields
\[ [3][1^2]=[4,1]\oplus[3,1^2], \]
so that the constituents of $[3][1^2][1^2]$ are obtained as follows:
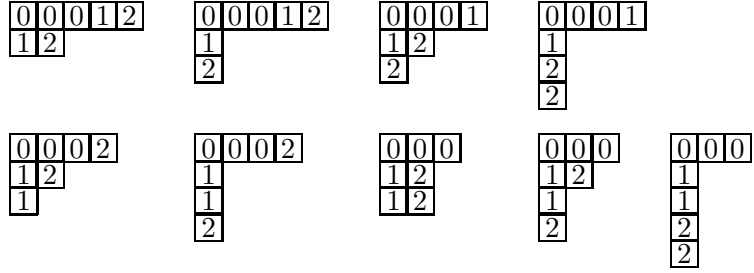
\[
\begin{picture}(330,120)(0,-102.5)
\put(0,-2.5){\framebox(10,10){\small$0$}}
\put(10,-2.5){\framebox(10,10){\small$0$}}
\put(20,-2.5){\framebox(10,10){\small$0$}}
\put(30,-2.5){\framebox(10,10){\small$1$}}
\put(40,-2.5){\framebox(10,10){\small$2$}}
\put(0,-12.5){\framebox(10,10){\small$1$}}
\put(10,-12.5){\framebox(10,10){\small$2$}}

\put(70,-2.5){\framebox(10,10){\small$0$}}
\put(80,-2.5){\framebox(10,10){\small$0$}}
\put(90,-2.5){\framebox(10,10){\small$0$}}
\put(100,-2.5){\framebox(10,10){\small$1$}}
\put(110,-2.5){\framebox(10,10){\small$2$}}
\put(70,-12.5){\framebox(10,10){\small$1$}}
\put(70,-22.5){\framebox(10,10){\small$2$}}

\put(140,-2.5){\framebox(10,10){\small$0$}}
\put(150,-2.5){\framebox(10,10){\small$0$}}
\put(160,-2.5){\framebox(10,10){\small$0$}}
\put(170,-2.5){\framebox(10,10){\small$1$}}
\put(140,-12.5){\framebox(10,10){\small$1$}}
\put(150,-12.5){\framebox(10,10){\small$2$}}
\put(140,-22.5){\framebox(10,10){\small$2$}}

\put(200,-2.5){\framebox(10,10){\small$0$}}
\put(210,-2.5){\framebox(10,10){\small$0$}}
\put(220,-2.5){\framebox(10,10){\small$0$}}
\put(230,-2.5){\framebox(10,10){\small$1$}}
\put(200,-12.5){\framebox(10,10){\small$1$}}
\put(200,-22.5){\framebox(10,10){\small$2$}}
\put(200,-32.5){\framebox(10,10){\small$2$}}

\put(0,-52.5){\framebox(10,10){\small$0$}}
\put(10,-52.5){\framebox(10,10){\small$0$}}
\put(20,-52.5){\framebox(10,10){\small$0$}}
\put(30,-52.5){\framebox(10,10){\small$2$}}
\put(0,-62.5){\framebox(10,10){\small$1$}}
\put(10,-62.5){\framebox(10,10){\small$2$}}
\put(0,-72.5){\framebox(10,10){\small$1$}}

\put(70,-52.5){\framebox(10,10){\small$0$}}
\put(80,-52.5){\framebox(10,10){\small$0$}}
\put(90,-52.5){\framebox(10,10){\small$0$}}
\put(100,-52.5){\framebox(10,10){\small$2$}}
\put(70,-62.5){\framebox(10,10){\small$1$}}
\put(70,-72.5){\framebox(10,10){\small$1$}}
\put(70,-82.5){\framebox(10,10){\small$2$}}

\put(140,-52.5){\framebox(10,10){\small$0$}}
\put(150,-52.5){\framebox(10,10){\small$0$}}
\put(160,-52.5){\framebox(10,10){\small$0$}}
\put(140,-62.5){\framebox(10,10){\small$1$}}
\put(150,-62.5){\framebox(10,10){\small$2$}}
\put(140,-72.5){\framebox(10,10){\small$1$}}
\put(150,-72.5){\framebox(10,10){\small$2$}}

\put(200,-52.5){\framebox(10,10){\small$0$}}
\put(210,-52.5){\framebox(10,10){\small$0$}}
\put(220,-52.5){\framebox(10,10){\small$0$}}
\put(200,-62.5){\framebox(10,10){\small$1$}}
\put(210,-62.5){\framebox(10,10){\small$2$}}
\put(200,-72.5){\framebox(10,10){\small$1$}}
\put(200,-82.5){\framebox(10,10){\small$2$}}

\put(250,-52.5){\framebox(10,10){\small$0$}}
\put(260,-52.5){\framebox(10,10){\small$0$}}
\put(270,-52.5){\framebox(10,10){\small$0$}}
\put(250,-62.5){\framebox(10,10){\small$1$}}
\put(250,-72.5){\framebox(10,10){\small$1$}}
\put(250,-82.5){\framebox(10,10){\small$2$}}
\put(250,-92.5){\framebox(10,10){\small$2$}}
\end{picture}
\]

Therefore
\[\fl [3][1^2][1^2]=[5,2]\oplus[5,1^2]\oplus2[4,2,1]\oplus2[4,1^3]
\oplus[3,2^2]\oplus[3,2,1^2]\oplus[3,1^4]. \]

This algorithm leads to the same diagrams as proposition 3,
if we replace the numbers $1$, ($2$) by the
symbols $\uparrow$, ($\downarrow$) respectively.

Proposition 3 describes how to obtain a basis of the subspace
$e^\alpha \mathcal{H}$ for a given diagram $[\alpha]$. The procedure described
above shows only that proposition 3 leads to the right dimension of
$e^\alpha \mathcal{H}$. But we have also seen that this dimension is never
higher than two. Thus it is easy to verify in every case that the
corresponding states $e^\alpha |\phi\rangle$ are linearly independent.

\section{Proof of (\ref{lemma})}

We first note, that $\sum_{i<j}\rho(ij)$ commutes with every
$\rho(\pi),\ \pi\in S_L$. Thus
\[ \sum_{i<j}\rho(ij)e^\alpha=\sum_{i<j}\mathcal{C}^\alpha \rho(ij)
\mathcal{R}^\alpha \]
Clearly $\mathcal{C}^\alpha \rho(ij)\mathcal{R}^\alpha=e^\alpha$
if $(ij)\in R^\alpha$ and
$\mathcal{C}^\alpha \rho(ij)\mathcal{R}^\alpha=-e^\alpha$ if $(ij)\in
C^\alpha$.
In the remaining case
$(ij)\not\in R^\alpha\cup C^\alpha$, there exists one site
$k\neq i,j$, which is in the same column as $i$ and in the same row as $j$.
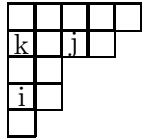
\[
\begin{picture}(50,70)(0,-52.5)
\put(0,-2.5){\framebox(10,10){}}
\put(10,-2.5){\framebox(10,10){}}
\put(20,-2.5){\framebox(10,10){}}
\put(30,-2.5){\framebox(10,10){}}
\put(40,-2.5){\framebox(10,10){}}
\put(0,-12.5){\framebox(10,10){\small k}}
\put(10,-12.5){\framebox(10,10){}}
\put(20,-12.5){\framebox(10,10){\small j}}
\put(30,-12.5){\framebox(10,10){}}
\put(0,-22.5){\framebox(10,10){}}
\put(10,-22.5){\framebox(10,10){}}
\put(0,-32.5){\framebox(10,10){\small i}}
\put(10,-32.5){\framebox(10,10){}}
\put(0,-42.5){\framebox(10,10){}}
\end{picture}
\]
As $(ij)=(ik)(ij)(kj)$, we find
\[\mathcal{C}^\alpha \rho(ij)\mathcal{R}^\alpha=
 \mathcal{C}^\alpha\rho(ik) \rho(ij)\rho(kj)\mathcal{R}^\alpha=-
\mathcal{C}^\alpha \rho(ij)\mathcal{R}^\alpha \]
and therefore
\[ \mathcal{C}^\alpha \rho(ij)\mathcal{R}^\alpha=0. \]

This proofs equation (\ref{lemma}).

\Figures
\begin{figure}
\caption{Spectrum for even number of electrons $N$}\label{sp1}
\end{figure}
\begin{figure}
\caption{Spectrum for odd number of electrons $N$}\label{sp2}
\end{figure}


\section*{References}
\begin{thebibliography}{99}

\bibitem{Dagotto} For a general review on the $t-J$ model and its relation
to high-temperature superconductors see Dagotto E 1994 {\it Rev. Mod. Phys.}
{\bf 66} 763; presently the $t-J$ ladders are intensively studied, see
Rice T M 1997 {\it Z. Phys. B} {\bf 103} 165.

\bibitem{Schlottmann} Schlottmann P 1987 {\it Phys. Rev. B} {\bf 36} 5177;
Sarkar S 199 {\it J. Phys. A} {\bf 24} 1137; Bares P A, Blatter G and
Ogata M 1991 {\it Phys. Rev. B} {\bf 44} 130; Essler F H L and
Korepin V E 1992 {\it Phys. Rev. B} {\bf 46} 9147.

\bibitem{Lai} Lai C K 1974 {\it J. Math. Phys.} {\bf 15} 1675; Sutherland B
1975 {\it Phys. Rev. B} {\bf 12} 3795.

\bibitem{Ogata} Ogata M and Shiba H 1990 {\it Phys. Rev. B} {\bf 41} 2326;
Penc K, Hallberg K, Mila F and Shiba H 1997 {\it Phys. Rev. B} {\bf 55}
15475.

\bibitem{Patterson} Patterson J D 1972 {\it Phys. Rev. B} {\bf 6} 1041;
van Dongen P and Vollhardt D 1989 {\it Phys. Rev. B} {\bf 40} 7252.

\bibitem{vDongen} Verg\'es J A, Guinea F, Gal\'an J, van Dongen P G J ,
Chiappe G and Louis E 1994 \PR {\bf B49} 15400.

\bibitem{LiMattis} Li C B, Mattis D C 1997 {\it Mod. Phys. Lett. B} {\bf 11}
115; Mattis D C , {\it ibidem} 123.

\bibitem{Kirson} Kirson M W 1997 {\it Phys. Rev. Lett.} {\bf 78} 4241.

\bibitem{Binz} Binz B 1997 {\it diploma thesis}, University of Fribourg,
unpublished.

\bibitem{Messiah} Messiah A 1961 {\it Quantum mechanics} volume II,
Appendix D (North-Holland Publishing Company).

\bibitem{Schensted} Schensted V 1976 {\it A course on the application
of group theory to quantum mechanics} (Neo Press, USA).

\bibitem{JamesKerber} James G D, Kerber A 1981 {\it The representation
theory of the symmetric group}, Encyclopedia of Mathematics
and its Applications, volume 16 (Addison-Wesley).


\bibitem{Tasaki} Tasaki H 1989 \PR  {\bf B40} 9192.

\bibitem{Mielke} Mielke A 1993 \PL  {\bf A174} 433.

\bibitem{Pieri} Pieri P 1997 {\it Mod. Phys. Lett. B} {\bf 10} 1277.

\bibitem{Auerbach} Auerbach A 1994 {\it Interacting electrons and quantum
magnetism} (Springer).

\end{thebibliography}
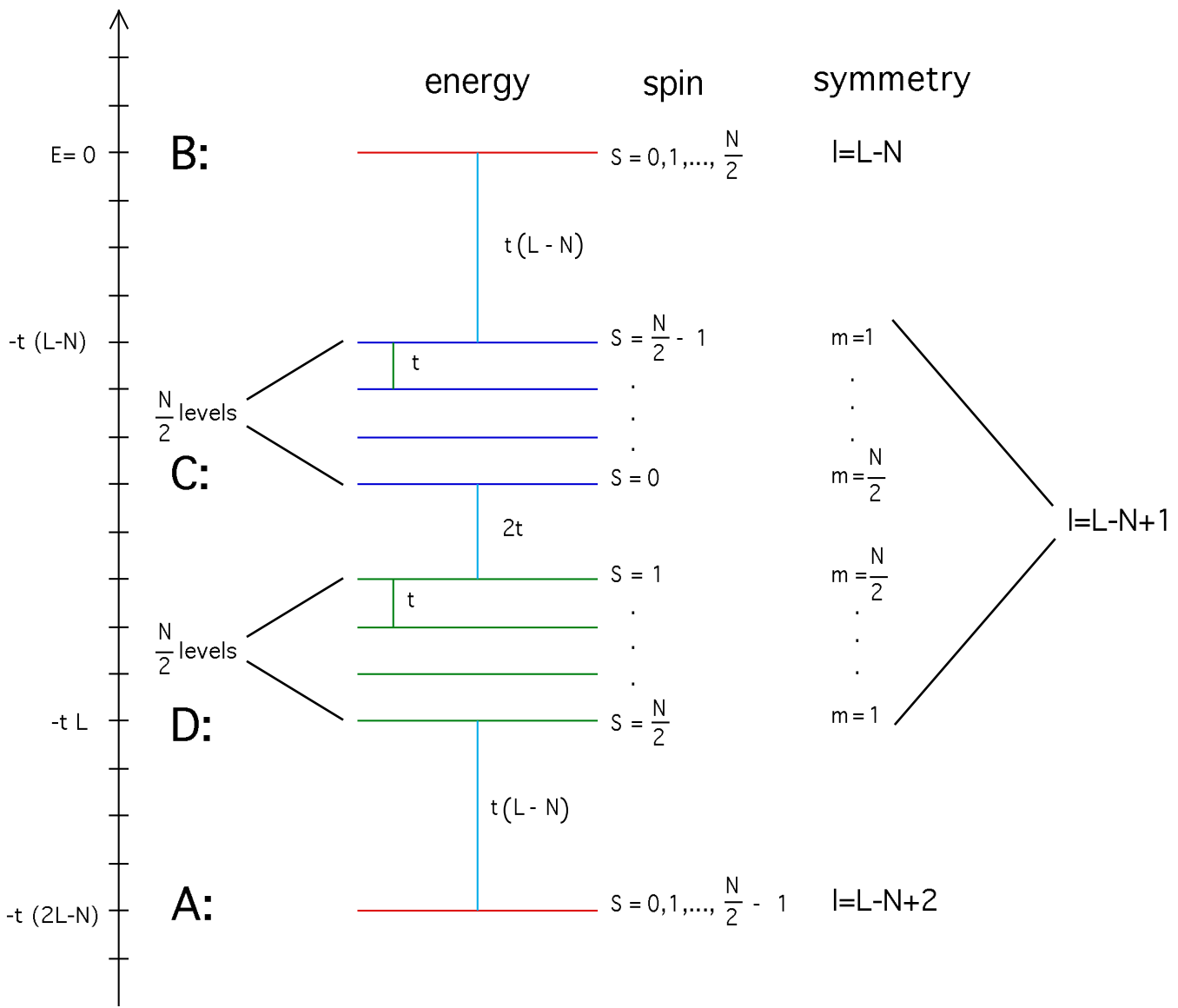
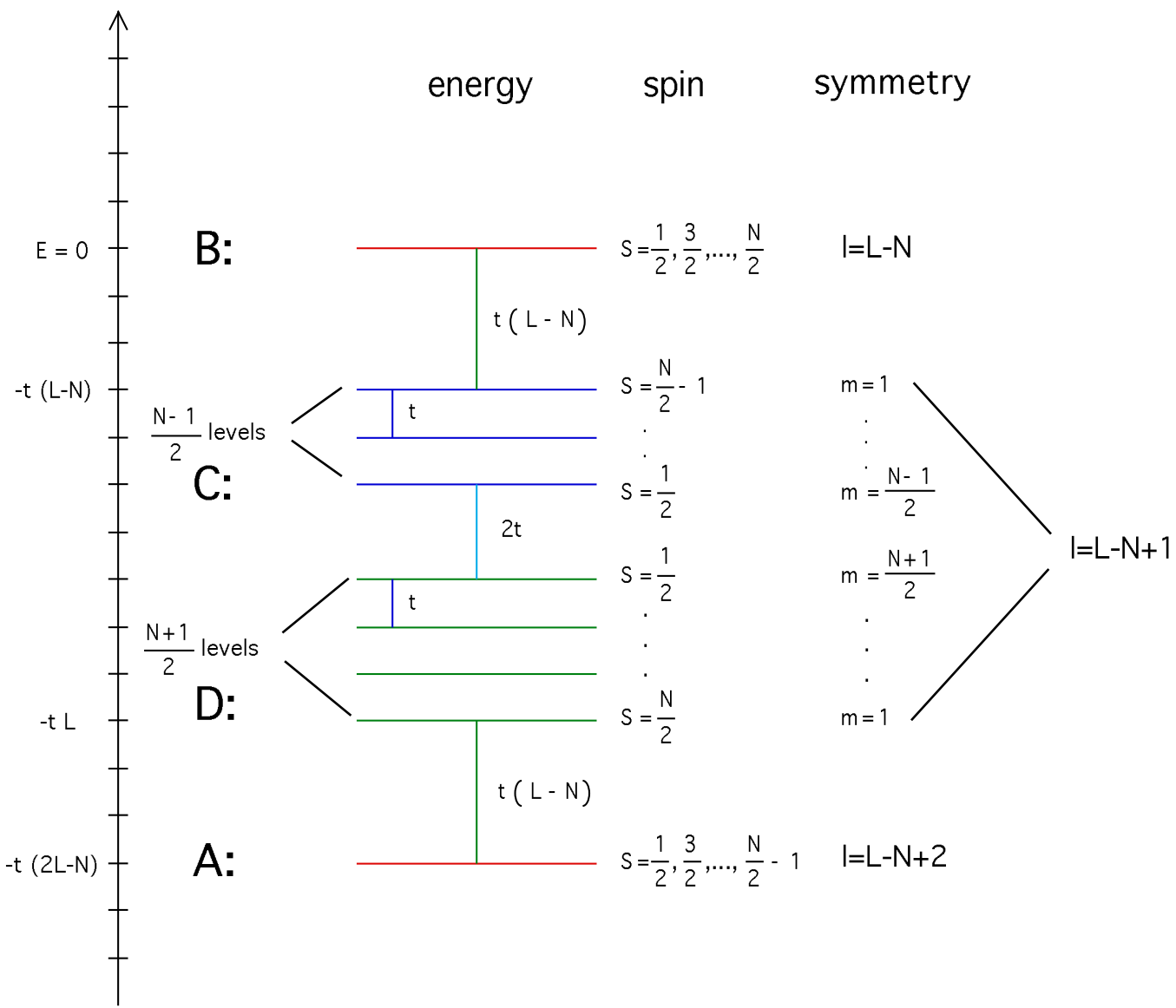
\end{document}